\documentclass[preprint,preprintnumbers]{revtex4}

\usepackage{graphicx}

\usepackage{amssymb}

\usepackage{mathrsfs}

\begin{document}

\title{Probing  Topological Superconductors  with Sound Waves }

\author{D. Schmeltzer}

\affiliation{Physics Department, City College of the City University of New York,  
New York, New York 10031, USA}



\begin{abstract} 
 
A new method is introduced for probing  Topological Superconductors.

The  integration of  the superconding  fermions  generates a topological $\mathbf{Chern-Simons}$  sound action . 

 Dislocations  induce   Majorana zero modes inside the sample, resulting in a new Hamiltonian   which couple  the Majorana modes, the electron field  and the non-Abelian strain  field sound. 
This Hamiltonian is used  to compute  the  anomalous sound  absorption. The Topological  superconductor absorbs sound below the  superconducting gap due to  the transition  between the quasi particles and  the Majorana fermions.  
  
The  sound waves   offers a new  tool for  detecting  Majorana fermions.


\end{abstract}

\maketitle
\textbf{I-Introduction}

\vspace  {0.2 in}

   Topological materials have been  discovered   recently  \cite{Kane,Zhang}.
A  new class of   insulators  coined Topological Insulators  are  characterized by the  second $\mathbf{Chern}$ number \cite{Nakahara}  which can  be measured  using  the Faraday and Kerr rotation \cite{Zhang}.
Similarly to Topological Insulators, Topological Superconductors have been found  and identified by the presence of  the Majorana zero modes  \cite{Read,Ivanov,Gurarie,Taylor,Alicea}. An indirect  observation   for the Majorana fermions  was obtained   from the differential  conductance measurement \cite{Mourik}.  
Since the  electrical   current  is not  conserved in  a superconductor,  the electromagnetic  field    can not be used to identify  the  Topological Superconductors. 
The response of a Topological Superconductor is  characterized  by a topological  $\mathbf{Chern-Simons}$ sound action as  we have for the Yang-Mills action \cite{Witten}.The coeficient of this action is the central charge  $\mathbf{c}$ which counts  the number  of the Majorana  edge  which are the signature for   the Topological Superconductor.
The  topological action contains high derivatives  and it is unlikely to be obseved in the laboratory.
For this reason we need to find an alternative method.

 Sound waves have played an important role in the field of Superconductivity. The first experimental measurement of the  superconducting  gap  was by ultrasound attenuation \cite{Bohm}.  A typical ultrasonic attenuation wave has a frequency of $10-100 MHZ$ which is too  small to break a pair and excite a quasi-particle,  since    gaps are  in the   range of  milli-electron volts. 
 This suggests that in order to enhance the sound absorption one needs to increase the density of the normal matter  $\rho_{n}$ in the superconductor. This can be done  with the help of   vortices  or dislocations. 

 In a superfluid  it has been  shown  that in the mixture   $^{4}He$-$^{3}He$,  the density of the normal matter $\rho_{n}$  increases  with $^{3}He$ \cite{Beamish,Kuklov}.  
The  technique  to measure  the normal density  $\rho_{n}$  is based on the   torsional oscillator  \cite{Kim}, the period of the oscillation is related to the moment of inertia  of the normal density.  Similar  measurements   have been done  by \cite{Murakawa}   to investigate     $^{3}H_{e}-B$ .

 Here we would like to propose a new method based  on  the   coupling  of sound waves to  Majorana fermions and electrons.  We will present a theory which demonstrates that the presence of the induced  Majorana modes affects the response  function . We compute   the  sound $\mathbf{polarization}$  and obtain the normalized sound  wave equation which is used to evaluate  the  shear  and longitudinal stress.   The  stress  serves as a probe for the Majorana fermions. 
The stress field  generates dislocations and  strain fields.  The effect of the  stress field on  the superconductor  can be measured using impedance techniques, by measuring the change of the quality factor $Q$ and resonance frequency of an ac-cut quartz transducer which   oscillates  in a shear or longitudinal mode

 Topological Superconductors   are characterized by a  winding number  and  the central charge $\mathbf{c}$ in the  the region where the chemical potential is positive.   At the interface between the   region of positive chemical potential  and a negative one, (where the superconductor is not a topological superconductor)    zero modes (half vortices)  appear  \cite{Alicea,Beenaker,Oreg}.   Half  vortices    are  identified with   the Majorana zero modes.  
The formation of the Majorana fermion is  due to the vanishing  of the Topological  Superconducting density, which on a two dimensional surface  is  identified  with the  vortices. Solids are characterized by dislocations which are either induced by large  external   deformations or  are a result of the crystal  growing process.  On a surface, the presence of the  dislocations  are similar to two dimensional vortices.  The  Majorana  zero modes are bound  to  the vortices on the surface of a Topological Superconductor.
This suggests that    dislocations   control the density  of the Majorana fermions.  By increasing the density of the vortices in a Superconductor, one can enhance   the acoustic  absorption.
We find an anomalous  absorption  for   frequencies   $\hat{\Delta}-\epsilon_{a} <\omega <\hat{\Delta}+\epsilon_{a} $, $\hat{\Delta}$ is the superconducting gap   and   $\epsilon_{a}$ represents the overlapping energies for two Majorana fermions. The anomalous absorption occurs  in   the forbidden  superconducting  frequency  region  $\omega<2\hat{\Delta} $.  The anomalous   absorption represents the finger print of the   Majorana fermions in a Topological Superconductor.

This results are obtained with  help of a  new  Hamiltonian. Using a space covariant transformation  induced by the sound waves we derive  the coupling between the non-Abelian strain  field,  the  Majorana fermion and the  electron field. 
 The sound waves "deforms" the  Topological Superconductor.  The deformation is obtained  with the help of      the  coordinates transformation  method \cite{Kleinert,Dislocations,Strain}.

 The plan of this paper is as follows . In Sec. $II$  we introduce the  Topological  Superconductor Hamiltonian. 
 Using  the $p-wave$ model we construct the deformation caused by elasticity, following  the methods given in the literature \cite{Kleinert,Nakahara,Dislocations,Strain}. We show that the  integration of  the fermions  generate the topological $\mathbf{Chern-Simons}$ term with the central charge which counts the Majorana modes on the boundary of the sample. 
In the second step we consider dislocations which induce Majorana zero modes inside the sample. For this situation  we obtain a new Hamiltonian  and  show that the Majorana modes couple to the electron field and sound. The details are given in $\textbf{Appendix -A}$ and  $\textbf{Appendix -B}$.   Sec. $III$ is  devoted  to the  computation of  the sound absorption, the details are given in $\textbf{Appendix -C}$. We compute  the  shear  and   longitudinal stress  and identify the anomalous sound absorption  which    confirms the presence  of   the Majorana modes. In section IV we present our conclusions.

\vspace{0.2 in}

\textbf{II- The    Topological Superconductor  in the presence of  an elastic strain field}

\vspace{0.2 in}

\noindent In order to demonstrate the anomalous absorption we consider  a  $p-wave$ superconductor \cite{Taylor,Read}. The $p-wave$ superconductor is   engineered using materials with strong  spin orbit interaction and   magnetic Zeeman fieldsi which  are in   proximity to   an  s-wave superconductors \cite{Oreg,Alicea} .
In   order to study the  stress response and absorption     we need to  know  the coupling between   the sound waves  and the superconductor.
The  form of  the electron phonon Hamiltonian  is problematic for spinors. The symmetry of the  coupling is  obtained from the coordinate transformation method \cite{Nakahara}.
 When a sound waves  excites  the crystal, the  crystal coordinates  are  modified. The external   strain field  modifies the position of the crystal atom.
\noindent  As a result  the $ p-wave$  is  replaced  by a  \textbf{deformed}  superconductor  Hamiltonian:
\begin{equation}
H=H^{(p-wave)}+\delta H^{(cr-p-wave)} +H^{(cr)}+H^{(ext.)}
\label{hamiltonian}
\end{equation}
\noindent  $H^{(p-wave)}$ is the model for the  $p-wave$  superconductor in the absence of the sound waves.
\begin{eqnarray}
&&H^{(p-wave)}=\frac{1}{2} \int\,d^2r C^{\dagger}(\vec{r},t)\Big[  \tau_{3}\Big(\frac{\hbar^2}{2m} (-i\vec{\partial}_{r})^2 -\mu_{F}(\vec{r})\Big)  -\Delta(\vec{r},t)\Big (\tau^{1}-i\tau^{2}\Big)(\partial_{1}+i\partial_{2})\nonumber\\&&+\Delta^{*}(\vec{r},t) \Big(\tau^{1}+i\tau^{2}\Big)(\partial_{1}-i\partial_{2}) \Big] C(\vec{r},t),
\nonumber\\&&
\end{eqnarray}
$\Delta (\vec{r},t)$ the pairing order field, $\mu_{F}(\vec{r})$ is the space dependent chemical potential  and $\tau^{1}$,$\tau^{2}$ $\tau^{3}$ are the Pauli matrices in the particle-hole space. We assume that in one region    $\mu_{F}(\vec{r})>0$   and in the complimentary   region $\mu_{F}(\vec{r})<0$.  For  $\mu_{F}(\vec{r})>0$ the superconductor is topological and is characterized by the  topological  invariant with the $\mathbf{Chern}$ number $\mathbf{Q}$$\mathbf{\in}$$\mathbf{Z}$ \cite{Alicea}, in the  region  $\mu_{F}(\vec{r})<0$ the superconductor  is non topological. At    the interface  $\mu_{F}(\vec{r})=0$ (between the two regions) the spectrum will contain  bound states,  Majorana zero modes. 
The change of sign of the chemical potential in space gives rise to the Majorana fermions.
In the literature a few proposal exist,  for the Majorana fermions, mostly at the interface   with a Ferromagnetic region \cite{Alicea} and vortices \cite{Jackiw}. Recently  new proposals  based on the   interplay of magnetism   and superconductivity  \cite{Schon,Yazdani} have been introduced. 
The change of sign of the chemical potential in a $p-wave$  superconductor can be  achieved  by the presence of  vortices  or dislocations (introduced by the growing process of the crystal)  or large  mechanical  stresses. The presence of the dislocations localized at position $\vec{R}_{b}$ gives rise to discrete  points  where the chemical potential $\mu_{F}(\vec{r}\approx  \vec{R}_{b})$  vanishes (see $\textbf{Appendix-A}$).

The crystal Hamiltonian  controls the sound propagation.  The  crystal Hamiltonian $H^{(cr)}$ is  given by:
\begin{eqnarray}
&&H^{(cr)}=\int\,d^2r\Big[\frac{\vec{\pi}^2}{2\rho}+\frac{\mu}{2}\Big(\partial_{i} u^{j}+\partial_{j}u^{i}\Big)^2+\frac{\lambda}{4}(\partial_{i}u^{i})^2\Big] ; i,j= 1,2\nonumber\\&&
\end{eqnarray} 
The solid Hamiltonian $H^{(cr)}$  contains the kinetic  and potential energy.   The kinetic energy  $\frac{\vec{\pi}^2}{2\rho}$  is determined by   the canonical momentum  $ \vec{\pi}$ and   mass density $\rho$. The potential  energy is characterized by the $Lame$ elastic constant $\lambda$ and shear modulus   $\mu$ \cite{Kleinert}.  

\noindent  $\vec{f}_{ext.}(\vec{q},t)=\vec{f}_{ex.}(\vec{q})e^{i\omega t}$  is the force of  an $A-C$  quartz transducer  which pumps energy    into the solid.   The external source   Hamiltonian    $H^{(ext.)}$ is  given by:
\begin{equation}
H^{(ext.)}=\int\,d^2r\vec{f}_{ext.}(\vec{r},t)\cdot\vec{u}(\vec{r},t)
\label{ext}
\end{equation}
    The sound propagation  modifies the  $p-wave$ .In  the presence of the sound waves, the coordinates  are modified   from $\vec{r}\equiv\vec{x}=[x^{(1},x^{(2)}]$ to $ \vec{x}+\vec{u}(\vec{x},t)=\vec{X}= \Big[X^{a=1}, X^{a=2}\Big]$  where   $\vec{u}(\vec{x},t)=[u^{(1)}(\vec{x},t),u^{(2)}(\vec{x},t)]$. The  sound modifies  the term  $\tau^{a}\partial_{a}$ (in the Hamiltonian Eq.$(2)$) to $\sum_{a=1,2}\tau^{a}\frac{\partial x^{(i)}}{\partial X^{(a)}}\nabla_{i}$.   $\frac{\partial x^{(i)}}{\partial X^{(a)}}$  is computed with the help of the  transformation    $\vec{X}(\vec{x})$  induced by the sound. As a result  a non-Abelian  strain field  $\vec{A}(\vec{x},t)$ is generated.  The non-Abelian   field  $\vec{A}(\vec{x},t)$ is written in terms of the  elastic strain field.The non-Abelian form is due to the  particle-hole operators $\tau^{1}$,$\tau^{2}$ $\tau^{3}$
 represents the effect of the sound waves which  is written  in terms    of   the elastic  field $\vec{u}(\vec{x},t)$. The full derivation of $H^{(deformed-p-wave)}$  Hamiltonian is   given in \textbf{Apendix -B}. The integration of the fermions with Hamiltonian  $H^{(deformed-p-wave)}$ and  $\mu_{F}>0$  generate a topological  sound action, 
\begin{equation}
S^{top-sound}=\frac{c}{96\pi}\int\,dt\int\,d ^2r\Big[\epsilon^{i,j,k}\omega_{i,a}(\partial_{j}\omega^{a}_{k}-\partial_{k}\omega^{a}_{j})+\frac{2}{3}\epsilon^{a,b,c}\omega_{i,a}\omega_{i,b}\omega_{i,c}\Big]
\label{top}
\end{equation} 
\noindent  $\mathbf{c}$  counts the number  of the Majorana   modes on  the edge of the sample for $\mu_{F}>0$. Due to the  high order derivatives  $S^{top-sound}$ is unlikely to be observed in the laboratory. 

We will consider a  chemical potential which changes sign  inside the sample due to vortices or dislocations. Due to vortices or dislocations we will have  additional  Majorana zero modes. 
We will focus  on the first non-trivial $\delta H^{(cr-p-wave)}\equiv H^{(deformed-p-wave)}-H^{(p-wave)}$ term which  contributes to  the absorption. Such a term  is linear in the Majorana  and in the  electron field.
\begin{eqnarray}
&&H^{(deformed-p-wave)}-H^{(p-wave)}\approx   \delta H^{(cr-p-wave)}=\int\,d^2r \Big[C^{\dagger}(\vec{r},t)\Big(\vec{A}(\vec{r},t)\cdot\vec{\partial}_{r}\Big) C(\vec{r},t)\Big],\nonumber\\&& 
\vec{A}(\vec{r},t)=I\vec{b}_{0}(\vec{r},t)+\tau_{1}\vec{b}_{1}(\vec{r},t)+\tau_{2}\vec{b}_{2}(\vec{r},t),\hspace{0.1 in}  \Delta=|\Delta|e^{i\alpha},\hspace{0.1 in}\vec{b}_{0}(\vec{r},t)=\partial_{t}\vec{u}(\vec{r},t)\nonumber\\&& 
 \vec{b}_{1}(\vec{r},t)=i|\Delta|\Big(\cos(\alpha)\partial_{2}\vec{u}(\vec{r},t)-\sin(\alpha)\partial_{1}\vec{u}(\vec{r},t)\Big)\hspace{0.1 in}
\vec{b}_{2}(\vec{r},t)=i|\Delta|\Big(\cos(\alpha)\partial_{1}\vec{u}(\vec{r},t)+\sin(\alpha)\partial_{2}\vec{u}(\vec{r},t)\Big) \nonumber\\&& 
\end{eqnarray}
The $ H^{(p-wave)}$ Hamiltonian is  diagonalized   with the help of the Bogoliubov - deGennes  transformation.  The  Hamiltonian has in addition to the positive eigenvalues also   Majorana zero modes \cite{Taylor,Read}.
 The  spinor  for non-zero energies are  given by $\Big[U(\vec{k}),V(\vec{k})\Big]^{T}$ and the zero modes spinor are  given by  $\Big[U_{b}(\vec{r},\phi),U^{*}_{b}(\vec{r},\phi)\Big]^T$ . The $Bogoliubov - deGennes$  Hamiltonian contains pairs of momentum $[\vec{k},-\vec{k}]$ and therefore the momentum integration is restricted to   half of the Brillouin zone. To cover the  entire Brillouin Zone we will  replace $\Big[U(\vec{k}),V(\vec{k})\Big]^{T}$ by a four component spinor   $\Phi(\vec{k})=\Big[U(\vec{k}),V(\vec{k}),U(-\vec{k}),V(-\vec{k})\Big]^{T}$, similarly for the Majorana spinor we will replace  $\Big[U_{b}(\vec{r},\phi),U^{*}_{b}(\vec{r},\phi)\Big]^T$   with a four component spinor 
$\hat{W}_{b}(\vec{r},\phi)=\Big[U_{b}(\vec{r},\phi),U^{*}_{b}(\vec{r},\phi),U_{b}(\vec{r},\phi),U^{*}_{b}(\vec{r},\phi)\Big]^T$ . 
\begin{eqnarray}
&&\mathbf{C}(\vec{r})=\int\,\frac{d^2k}{(2\pi)^2}e^{i\vec{k}\vec{r}}\Big[\eta(\vec{k})\Phi_(\vec{k})+\eta(-\vec{k})\Gamma
\Phi^{*}(\vec{k})\Big]+\sum_{b=1}^{2n}\gamma_{b}\hat{W}_{b}(\vec{r},\phi)\nonumber\\&&
\mathbf{C}^{\dagger}(\vec{r})=\int\,\frac{d^2k}{(2\pi)^2}e^{-i\vec{k}\vec{r}}\Big[\eta^{\dagger}(\vec{k})[\Phi^*(\vec{k})]^T+\eta(-\vec{k}) [\Phi(-\vec{k})]^{T}\Gamma
\Big]+\sum_{b=1}^{2n}\gamma_{b}[\hat{W}_{b}^*(\vec{r},\phi)]^T\nonumber\\&&
\end{eqnarray}
 $\mathbf{C}(\vec{r})$  and $\mathbf{C}^{\dagger}(\vec{r})$ are the quasi particles operator for the superconductor.  
The matrix $\Gamma$ ensures the (pseudo) reality conditions  \cite{Jackiw}. 
\begin{equation}
\Gamma \mathbf{C}^{\dagger}(\vec{r})=\mathbf{C}(\vec{r}), \hspace{0.1 in} \Gamma \Phi^{*}_{E}(\vec{k})=\Phi_{-E}(\vec{k})
\label{conju}
\end{equation}
The operators $\mathbf{C}(\vec{r})$ , $\mathbf{C}^{\dagger}(\vec{r})$   also contain the $\textbf{Majorana zero modes}$ $\sum_{b=1}^{2n}\gamma_{b}\hat{W}_{b}(\vec{r},\phi)$ , $\sum_{b=1}^{2n}\gamma_{b}[\hat{W}_{b}^*(\vec{r},\phi)]^T$ where $\gamma_{b}$  are the Majorana operators. 
Using the mode expansion given in Eq.$(7)$ we will replace  the Hamiltonian $\delta H^{(cr-p-wave)}$ by: 
\begin{eqnarray}
&&\delta H^{(cr-p-wave)}=\nonumber\\&&\int\,d^2 r\int\,\frac{d^2k}{(2\pi)^2}e^{-i\vec{k}\vec{r}}\Big[\eta^{\dagger}(\vec{k})[\Phi^*(\vec{k})]^T+\eta(-\vec{k}) [\Phi(-\vec{k})]^{T}\Gamma
\Big]\Big(\vec{A}(\vec{r},t)\cdot\vec{\partial}\Big)\sum_{b=1}^{2n}\gamma_{b}[\hat{W}_{b}^*(\vec{r},\phi)]^T\Big]+h.c.\nonumber\\&&
\end{eqnarray} 

\vspace{0.2 in}

\textbf{III- Computation of the anomalous sound absorption}

\vspace{0.2 in }

In order to perform the computation in Eq.$(9)$ we need to know the spatial wave function for the Majorana fermions.  Due  to their   localization in space and low density we can  assume a model of randomly distributed  Majorana fermions.
 We will consider a situation where the correlation  within the nearest neighbor  pair $\Big[\gamma_{b=2a-1}, \gamma_{b'=2a}\Big]$ is significant and negligible otherwise. 
This allows to introduce the   fermion operators  $ \zeta^{\dagger}_{a}$,$\zeta_{a}$ ;
\begin{equation}
\gamma_{2a-1}=\frac{1}{\sqrt{2}}\Big[\zeta^{\dagger}_{a}+\zeta_{a}\Big],\hspace{0.1 in} \gamma_{2a}=\frac{1}{i\sqrt{2}}\Big[\zeta^{\dagger}_{a}-\zeta_{a}\Big] , a=1.2..n
\label{zeta}
\end{equation}
The overlap between the pair   $\Big[\gamma_{b=2a-1}, \gamma_{b'=2a}\Big]$    is  given  by  the overlapping energy $\epsilon _{a}$.  The  Majorana Hamiltonian is given by  $ H^{Majorana}=\sum_{a=1}^{a=n}\epsilon_{a}\zeta^{\dagger}_{a}\zeta_{a}$.
The ground state with the Majorana fermions is given by $ |G,\epsilon_{a=1}...\epsilon_{a=n}\rangle$ . This ground state $|G,\epsilon_{a=1}...\epsilon_{a=n}\rangle$ annihilates  the operators  $\eta(\vec{k})$ and  $\zeta_{a}$,   $\eta(\vec{k})|G,\epsilon_{a=1}...\epsilon_{a=n}\rangle=0$,  $\zeta_{a}  |G,\epsilon_{a=1}...\epsilon_{a=n}\rangle=0$. 
We compute the expectation value with respect the ground state $ |G,\epsilon_{a=1}...\epsilon_{a=n}\rangle$ and  find the sound $\mathbf{ polarization}$  $\mathbf{\Pi}[\vec{q},\omega]$.  
\begin{equation}
\langle G,\epsilon_{a=1}...\epsilon_{a=n}| T e^{\frac{-i}{\hbar}\int_{-\infty}^{\infty}\,dt \delta H^{(cr-p-wave)}} |G,\epsilon_{a=1}...\epsilon_{a=n}\rangle \approx T e^{\frac{-i}{\hbar}\int_{-\infty}^{\infty}\,dt \delta H^{(cr)}[\partial_{i}u^{j}(\vec{r},t),\partial_{t}u^{i}(\vec{r},t)]}
\label{eal} 
\end{equation}
As a result the sound wave Hamiltonian is replaced  by:
$H=H^{(cr)}+ H^{(ext.)}+\delta H^{(cr)}$ where  $\delta H^{(cr)}$  is given by Eq.$(11)$. (It
gives the sound absorption induced by the transition between the superconductor   quasi-particles  with energy $E=\hbar v\sqrt{(\hat{\epsilon}(\vec{k})-\hat{\mu}_{F})^2+|\Delta|^2 k^2}$ and the Majorana fermions $ \epsilon_{a}$.)
We evaluate   Eq.$(11)$   to order $\frac{1}{\hbar^2}$ and obtain the  polarization diagram $\mathbf{\Pi}[\vec{q},\omega]$.  
\begin{equation}
\delta H^{(cr)}=\int\,\frac{d^2k}{(2\pi)^2}\int\,\frac{d\omega}{2\pi}\mathbf{\Pi}[\vec{q},\omega]\vec{u}(\vec{q},\omega)\cdot\vec{u}(-\vec{q},-\omega),\hspace{0.1 in} \mathbf{\Pi}[\vec{q},\omega]\equiv\mathbf{Re}\mathbf{\Pi}[\vec{q},\omega]+ \mathbf{Im}\mathbf{\Pi}[\vec{q},\omega]
\label{eq.}
\end{equation}
The   polarization   $\mathbf{\Pi}[\vec{q},\omega]$   is computed in  $\textbf{Appendix-C}$.
Using typical values for the sound waves  frequency  $\omega=10^{9}Hz$ and Fermi momentum  $k_{F}=10^{9}m^{-1} $ we find   that the  value of the self energy  
 $\mathbf{\Pi}[\vec{q},\omega]$ is in the range  $100\frac{Newton}{m^2}$,
therefore  the effect on the  phonon frequency is negligible,  and we   will consider  only the imaginary part  of the polarization  $\mathbf{\Pi}[\vec{q},\omega]$  which determines the  absorption.
\begin{eqnarray}
&&\mathbf{\Pi}[\vec{q},\omega]= \frac{ \hbar \omega^2 k^{3}_{F}}{8\pi}\Big[1- 2|\Delta|\frac{|\vec{q}|}{|\omega|}\cos[\alpha-\beta(\vec{q})]-4|\Delta|^2(\frac{|\vec{q}|}{|\omega|})^2\Big]\cdot \sum_{a=1}^{n}\nonumber\\&&\int\,\frac{d\hat{E}}{\sqrt{\hat{E}^2-\Delta^2 k^{2}_{F}}}
 \frac{(\sqrt{\hat{E}^2-\Delta^2 k^{2}_{F}}-\hat{E})^2}{(\sqrt{\hat{E}^2-\Delta^2 k^{2}_{F}}-\hat{E})^2+\Delta^2 k^{2}_{F}}\Theta[\hat{E}-\Delta k_{F}]\cdot\nonumber\\&&
\Big[\Theta[\omega]\Big((\Theta[\hat{E}]\Theta[\hat{\epsilon}_{a}]-
\Theta[-\hat{E}]\Theta[-\hat{\epsilon}_{a}])\delta[\omega-(\hat{E}+\hat{\epsilon}_{a})]
+(\Theta[-\hat{E}]\Theta[\hat{\epsilon}_{a}]+
\Theta[\hat{E}]\Theta[-\hat{\epsilon}_{a}])\delta[\omega-(\hat{E}-\hat{\epsilon}_{a})]\Big)+\nonumber\\&&
\Big[\Theta[-\omega]\Big((\Theta[\hat{E}]\Theta[\hat{\epsilon}_{a}]-
\Theta[-\hat{E}]\Theta[-\hat{\epsilon}_{a}])\delta[\omega+(\hat{E}+\hat{\epsilon}_{a})] +(\Theta[\hat{E}]\Theta[-\hat{\epsilon}_{a}]+\Theta[\hat{E}]\Theta[-\hat{\epsilon}_{a}])\delta[\omega+(\hat{E}-\hat{\epsilon}_{a})]\Big)\Big]\nonumber\\&&
\end{eqnarray}
In   Eq.$(12)$  we  used the definitions $\epsilon(\vec{k})=\frac{\hbar^2}{2m}\vec{k}^2$, $\hat{\epsilon}(\vec{k})=\frac{\epsilon(\vec{k})}{\hbar v}$,$\hat{\mu}_{F}=\frac{\mu_{F}}{\hbar v}$,
 $\hat{E}=\frac{E}{\hbar v}=\sqrt{(\hat{\epsilon}(\vec{k})-\hat{\mu}_{F})^2+|\Delta|^2 k^2}$.

\noindent The first line of Eq.$(13)$ is determined by the structure of the sound field $\vec{A}(\vec{r},t)$  given in Eq.$(3)$. $\sum_{a=1}^{n}$  represents the sum over the uncorrelated Majorana fermions $\epsilon_{a}$.The second  line of Eq.$(13)$ contains the  density of states  $\frac{\Theta[\hat{E}-\Delta k_{F}]}{\sqrt{\hat{E}^2-\Delta^2 k^{2}_{F}}}$ and the coherence factor   $\frac{(\sqrt{\hat{E}^2-\Delta^2 k^{2}_{F}}-\hat{E})^2}{2((\sqrt{\hat{E}^2-\Delta^2 k^{2}_{F}}-\hat{E})^2+\Delta^2 k^{2}_{F})}$ ( introduced by the spinors 
$\Phi(\vec{k})$, $\hat{W}_{a}(\vec{r},\phi)$)

\noindent The  $\mathbf{third}$ line  with $\Theta[\omega]$ ($\omega>0$) describes the sound absorption.

\noindent  The $\mathbf{first}$ term in the  the third line describes the  creation of a Majorana  hole  with the  energy $\epsilon_{a}$ (destruction of Majorana particle with the  energy $-\epsilon_{a}$) and the creation of a electron with  energy $E$. As a result the absorption of sound  will occur for frequencies $\omega>\hat{E}+\hat{\epsilon}_{a}$ with the threshold   absorption  $\omega>|\Delta| k_{F}+\hat{\epsilon}_{a}$. In the absence of the Majorana term the absorption will occur $\omega>2\hat{ E}$,resulting in a forbidden frequency  region   $\omega>2|\Delta| k_{F}$.
 
\noindent The $\mathbf{second}$ term in the third line describes a situation where the Majorana states are occupied at the energy $ \hat{\epsilon}_{a}$, and the sound   absorption will excite the electron to the state  energy $\hat{E}$. As a result the sound will be absorbed for frequencies  $\omega>\hat{E}-\hat{\epsilon}_{a}$ ,resulting in the threshold absorption $\omega>|\Delta| k_{F}-\hat{\epsilon}_{a}$. This absorption will be controlled by the Fermi Dirac  occupation function which  at finite temperatures $T$ replaces $ \Theta[-\epsilon_{a}]$  with  $n_{F}[\epsilon_{a}]=\frac{1}{e^{\frac{\epsilon_{a}}{K_{B}T}}+1}$. The absorption region $ |\Delta| k_{F}-\hat{\epsilon}_{a}<\omega<|\Delta| k_{F}+\hat{\epsilon}_{a}$ will be called the \textbf{anomalous absorption}. Due to the fact that $ |\Delta| k_{F}>>\hat{\epsilon}_{a}$ the presence of the Fermi Dirac function $n_{F}[\epsilon_{a}]$ in the absorption formula gives the possibility to determine the energy distributions  of the Majorana fermions.

\noindent The  $\mathbf{fourth}$  line with $\Theta[-\omega]$ ($\omega<0$) describes the sound emission.

 We will use the polarization operator   $\mathbf{\Pi}[\vec{q},\omega]$ obtained in Eq.$(13)$ to compute the sound waves propagation in the crystal.
We note that due to  the random distribution of the Majorana fermions the  sound polarization  is isotropic and will affect the sound waves equations  in a symmetric way (see Eq.$(14)$). The   crystal Hamiltonian is  normalized  by the  sound polarization    $\mathbf{\Pi}[\vec{q},\omega]\vec{u}(\vec{q},\omega)\cdot\vec{u}(-\vec{q},-\omega)$. The sound  wave   
equation  for   the external force $f^{(i)}(\vec{q},\omega)$ is given by:
\begin{equation}
\Big[(-\omega^2+\mu q^{2}+\mathbf{\Pi}[\vec{q},\omega])\delta _{i,j} +(\lambda+\mu +\frac{\mathbf{\Pi}[\vec{q},\omega]}{q^2})q_{i}q_{j}\Big]u^{(j)}(\vec{q},\omega)=f^{(i)}(\vec{q},\omega) ;\hspace{0.1 in} i,j=1,2
\label{eqm}
\end{equation}
We consider a situation where the force   in  the $x$ direction is zero  $ f^{(1)}(\vec{q},\omega)=0$, and the force in the $y$  direction is $ f^{(2)}(\vec{q},\omega)\neq0$. We compute    the \textbf{shear stress}   $\mathbf{\sigma_{1,2}}[\vec{q},\omega]$, the \textbf{longitudinal stress} $\mathbf{\sigma_{2,2}}[\vec{q},\omega]$,  the \textbf{strain tensor} $\mathbf {\epsilon}_{1,2}[\vec{q},\omega]$,  $\mathbf {\epsilon}_{2,2}[\vec{q},\omega]$  the  impedance  $\mathbf{Z}[\vec{q},\omega]$ in terms  of the $ \textbf{shear modulus}$ $\mu$ and $\textbf{Lame}$ elastic constant  $\lambda$ .Using the solutions obtained from Eq.$(14)$ we find:
\begin{eqnarray}
&& \mathbf{\sigma_{1,2}}[\vec{q},\omega]=\epsilon_{1,2}[\vec{q},\omega] 2\mu  ,\hspace{0.1 in}\epsilon_{1,2}[\vec{q},\omega]\equiv\frac{1}{2}\Big[i q_{2}u^{(1)}(\vec{q},\omega)+i q_{1}u^{(2)}(\vec{q},\omega)\Big] \nonumber\\&&
\mathbf{\sigma_{2,2}}[\vec{q},\omega]=\epsilon_{2,2}[\vec{q},\omega] (3\lambda +2\mu)  ,\hspace{0.1 in}\epsilon_{2,2}[\vec{q},\omega]\equiv\Big[i q_{2}u^{(2)}(\vec{q},\omega)\Big]\nonumber\\&&
\mathbf{Z}_{1,2}[\vec{q},\omega]= \frac{\mathbf{\sigma_{1,2}}[\vec{q},\omega]}{f^{(2)}[\vec{q},\omega]}\equiv\mathbf{R}_{1,2}[\vec{q},\omega] +i\mathbf{X}_{1,2}[\vec{q},\omega]=\frac{ i 2\mu\frac{ q^3_{1}}{q^2}}{\mu q^2+i\mathbf{Im}\mathbf{\Pi}[\vec{q},\omega]}\nonumber\\&&
\mathbf{Z}_{2,2}[\vec{q},\omega]= \frac{\mathbf{\sigma_{2,2}}[\vec{q},\omega]}{f^{(2)}[\vec{q},\omega]}\equiv\mathbf{R}_{2,2}[\vec{q},\omega] +i\mathbf{X}_{2,2}[\vec{q},\omega]=\frac{ i(3\lambda +2\mu)  q_{2}}{\mu q^2+i\mathbf{Im}\mathbf{\Pi}[\vec{q},\omega]}\nonumber\\&&
\end{eqnarray}
 We plot  the real part of the impedance   $ \mathbf{R}_{1,2}[\vec{q},\omega]$  for a fixed Majorana energy, ignoring the energy dispersion   of  the Majorana  fermions. Due to the isotropic form  of the polarization the longitudinal   impedance   $ \mathbf{R}_{2,2}[\vec{q},\omega]$ has the same frequency dependence as   $\mathbf{R}_{1,2}[\vec{q},\omega]$ (see Eq.$(14)$). We will use in the plot   a narrow  energies distribution  $\frac{\epsilon_{a}}{\omega_{F}}\approx  0.05$. 

\noindent The  $\mathbf{thin}$  line gives the  anomalous absorption   $\frac{ |\Delta| k_{F}-\epsilon_{a}}{\omega_{F}}<\frac{\omega}{\omega_{F}}< \frac{ |\Delta| k_{F}+\epsilon_{a}}{\omega_{F}}$  for the   Majorana energy $\frac{ \epsilon_{a} }{\omega_{F}}\approx 0.05$  and  the  superconducting  gap   $\frac{ |\Delta| k_{F}}{\omega_{F}}\approx 0.25$.  $\omega_{F}= \frac{\mu_{F}}{\hbar}$  is the corresponding frequency  for  the  Fermi  energy.

\noindent The $\mathbf{ thick}$ line shows the  absorption in the  $\mathbf{absence}$ of the Majorana   fermions  for frequencies   $\frac{ \omega }{\omega_{F}} <\frac{2\mathbf{|\Delta|k_{F}}}{\omega_{F}}$  (The  form of the gap $\Delta |\vec{k}| $  is a result of the Hamiltonian given in   Eq.$(2)$.)
\begin{figure}
\begin{center}
\includegraphics[width=5.0 in ]{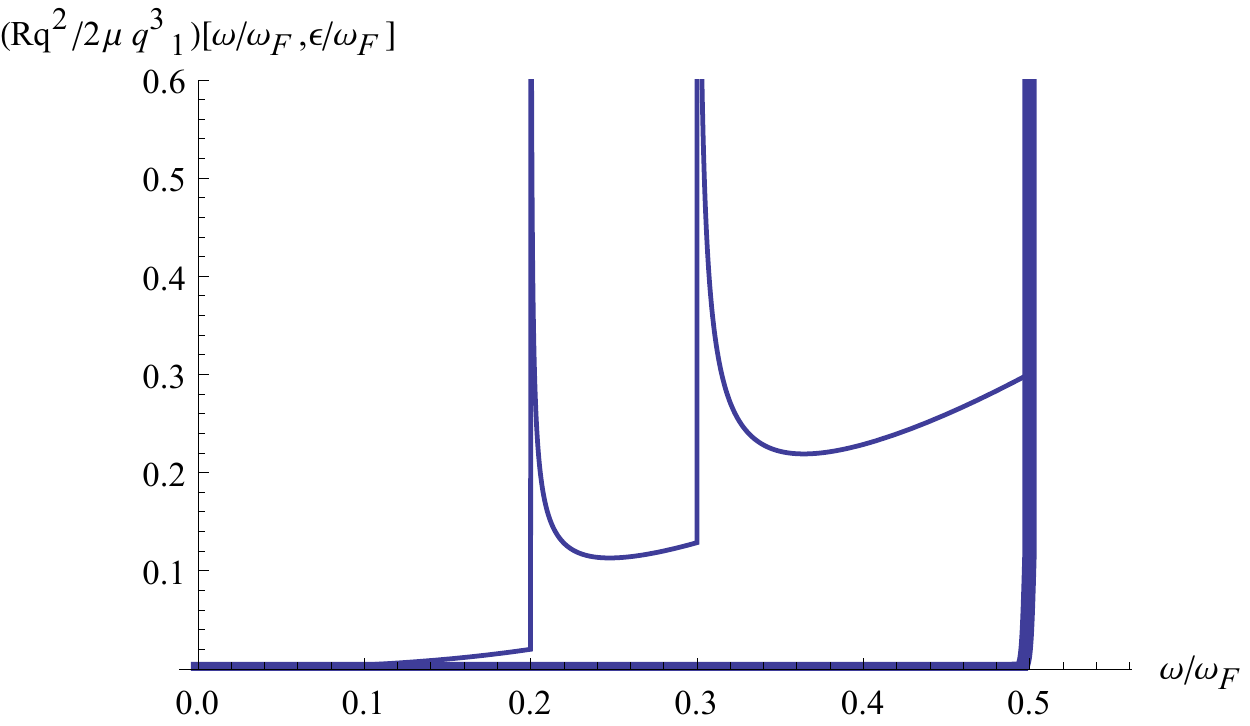}
\end{center}
\caption{The real part of the impedance   is $ \mathbf{R}[\vec{q},\omega]$ .
 The  $\mathbf{thin}$  line gives the  anomalous absorption   $\frac{ |\Delta| k_{F}-\epsilon_{a}}{\omega_{F}}<\frac{\omega}{\omega_{F}}< \frac{ |\Delta| k_{F}+\epsilon_{a}}{\omega_{F}}$  for the   Majorana energy $\frac{ \epsilon_{a} }{\omega_{F}}\approx 0.05$  and  the  superconducting  gap   $\frac{ |\Delta| k_{F}}{\omega_{F}}\approx 0.25$. The anomalous absorption is plotted for low temperatures   such that $n_{F}[\epsilon_{a}]\approx 1$. The $\mathbf{thick}$ line shows the sound absorption in the absence of the Majorana, the absorption $\mathbf{starts}$   for  frequencies  $\frac{ \omega }{\omega_{F}} >\frac{2\mathbf{|\Delta|k_{F}}}{\omega_{F}}=0.5$ }
\end{figure}
\noindent Since disorder is  present, it raises the question how to differentiate the absorption  between the Majorana fermions  and  charged impurities.
Charged impurities will give rise to absorption for frequencies $\omega>\hat{E}- \hat{\epsilon}_{imp.}$    and temperatures $ K_{B}T> 2|\Delta| k_{F}$, contrary to the absorption induced by the  Majorana fermions  at $T\rightarrow 0$.

\vspace{0.2 in}

\textbf{IV-Conclusions} 

\vspace{0.2 in}

\noindent To conclude a new proposal  for probing superconductors  with sound waves has been introduced.
An anomalous absorption is found in the frequency region $|\Delta| k_{F}-\hat{\epsilon}_{a}<\omega<|\Delta|k_{F}+\hat{\epsilon}_{a}$  and the energy distributions  of the Majorana fermions  can be determined. This gives  direct evidence  for  the presence of the Majorana fermions. The Majorana fermions are the finger print of   the Topological Superconductors. Therefore by measuring  the anomalous absorption we have  identified a method for  detecting Majorana fermions  and probing  the response of  the Topological  Superconductors.

\vspace{0.2 in}

\textbf{Appendix -A}

\vspace{0.2 in}

Due to the large deformation introduced by the dislocation localized at $\vec{R}_{b}$,  the crystal coordinates  are modified from $\vec{r}=[x,y]$ to  $\vec{R}(\vec{r})=[X(\vec{r}),Y(\vec{r})]$.This allows to introduce the static strain field $ e^{a}_{i}=\partial_{i} X^{a}$, $a=1,2$ and  $i=1,2$.
Due to the dislocation the  the area $dX dY$ is modified to $J[\vec{R}_{b}]dx dy$ where  the $ J[\vec{R}_{b}]$ is the Jacobian of the transformation, As a result the chemical potential $\mu_{F}$  becomes  $\mu_{F}\rightarrow \mu_{F}(\vec{r})\equiv\mu_{F}J[\vec{R}_{b}]$.
For an edge dislocation  the coordinate transformation  is given by :
$X=x$ and $Y=y-\frac{B^{(2)}}{2\pi}tan^{-1}\frac{y-R^{(y)}_{b}}{x-R^{(x)}_{b}}$. 
$B^{(2)}$ is  the Burgers vector in the $y$ direction. This transformation   gives for the Jacobian $J[\vec{R}_{b}]\equiv \Big(e^{ 1}_{1}e^{ 2}_{2}- e^{ 1}_{2}e^{ 2}_{1}\Big) =1- \frac{B^{(2)}}{2\pi}\frac{x-R^{(x)}_{b}}{(x-R^{(x)}_{b})^2+(y-R^{(y)}_{b})^2}$ \cite{Kleinert}.  As a result the  chemical   potential   $\mu_{F}(\vec{r}\approx \vec{R}_{b})$ vanishes.

\vspace{0.2 in}

\textbf{Appendix- B}

\vspace{0.2 in}

\noindent  The sound waves field $\vec{u}(\vec{x},t)$  change  the coordinates     from $\vec{r}\equiv \vec{x}=[x,y]$ to $ \vec{x}+\vec{u}(\vec{x},t)=\vec{X}= \Big[X^{a=1}, X^{a=2}\Big]$  where   $\vec{u}(\vec{x},t)=[u^{(1)}(\vec{x},t),u^{(2)}(\vec{x},t)]$. The strain field are  defined by $E^{i}_{a}=\partial_{a} x^{i}$, $e^{a}_{i}=\partial_{i}X^{a}$ are   related. We have:
 $\sum_{i=1}^{2}e^{a}_{i}E^{i}_{b}=\delta_{a,b}$,   $\sum_{a=1,2}e^{a}_{i}e^{a}_{j}= g_{i,j}$ and $E_{i,b}=g_{i,j}E^{j}_{b}$.
For the time component we have $E^{t}_{i}=e^{t}_{i} \delta_{i,t}$.

\noindent The derivatives transform like vectors, $\partial_{a}=\partial_{a} u^{1}(\vec{x},t)\partial_{1}+ \partial_{a} u^{2}(\vec{x},t)\partial_{2}$.  For   the sound waves   $\vec{u}(\vec{r},t)$ we have:   $e^{1}_{i}=\delta_{1,i}-\partial_{i}u^{1}(\vec{x},t)$, $e^{2}_{i}=\delta_{2,i}-\partial_{i}u^{2}(\vec{x},t)$,   $e^{a}_{t}=-\partial_{t}u^{a}(\vec{x},t)$ for $ a=1,2$.
The integration area element $d^2x$ in   Eq.$(2)$ is  multiplied by the Jacobian $J=  \Big[e^{1}_{1}e^{2}_{1}- e^{1}_{2}e^{2}_{1}\Big]\equiv Det[e^{a}_{i}]$.   The deformed $p-wave$  Hamiltonian $H^{(deformed-p-wave)}$ takes the form:
\begin{eqnarray}
&&H^{(deformed-p-wave)}= \frac{1}{2}\int\,d^2x Det[e^{a}_{i}] C^{\dagger}(\vec{r},t)\Big[  \tau^{3}\Big(\frac{-\hbar^2}{2m} (\sum_{a=1}^{2}E^{i}_{a}E^{j}_{a} \nabla_{i}  \nabla_{j}) -\mu_{F}(\vec{x})\Big)  -\Delta(\vec{x},t)\Big (\tau^{1}-i\tau^{2}\Big)(E^{i}_{1}\nabla_{i}+iE^{i}_{2}\nabla_{i})\nonumber\\&&+\Delta^{*}(\vec{x},t) \Big(\tau^{1}+i\tau^{2}\Big)(E^{i}_{1}\nabla_{i}-iE^{i}_{2}\nabla_{i}) \Big] C(\vec{x},t)\nonumber\\&&
\end{eqnarray}
\noindent $\mathbf{\nabla}_{i}$ is  the covariant derivative  given in terms of the spin connection:

\noindent $\mathbf{\omega}_{i}^{a,b}[\tau^{a},\tau^{b}]$, $\mathbf{\nabla}_{i}\equiv \partial_{i}+\frac{1}{8}\mathbf{\omega}_{i}^{a,b}[\tau^{a},\tau^{b}]$.  The spin connection has been derived in \cite{Dislocations} in terms of  $E^{i}_{a}$ and $e^{a}_{i}$. The  spin connection  is determined  from the zero torsion condition, $\nabla_{i}e^{a}_{j}- \nabla_{j}e^{a}_{i}=0 $. We have,
\begin{equation}
\omega^{a}_{i}=\epsilon^{a,b,c}E^{j}_{c}(\partial_{i}E_{j,b}-
\partial_{j}E_{i,b})-\frac{1}{2}\epsilon^{b,c,d}(E^{j}_{c} E^{k}_{c}\partial_{k} E_{j,d})e^{a}_{i}
\label{notorsion}
\end{equation}
Where $E_{i,b}=g_{i,j}E^{j}_{b}$.
 
Once the spin connection is know we can perform the path integral over the Dirac' fermion. As for the Yang-Mills theory  the  fermion integration in $2+1$ dimensions generate  a  non-Abelian Chern-Simons  term.  
We perform the path integral   integration  tor the fermion field  $C^{\dagger}(\vec{r},t)$  and  $C(\vec{r},t)$  and obtain the effective sound action $S^{top-sound}$  \cite{Strain}.    The topological term is given by, 
\begin{equation}
S^{top-sound}=\frac{c}{96\pi}\int\,dt\int\,d ^2r\Big[\epsilon^{i,j,k}\omega_{i,a}(\partial_{j}\omega^{a}_{k}-\partial_{k}\omega^{a}_{j})+\frac{2}{3}\epsilon^{a,b,c}\omega_{i,a}\omega_{i,b}\omega_{i,c}\Big]
\label{topolo}
\end{equation}

\noindent  where $\omega_{i,c}=g_{i,j}\omega^{j}_{c}$.
$\mathbf{c}$ counts the number  of the edge  modes. For the Topological Superconductor we  have  $\mu_{F}>0$ and $\mathbf{c}$ is non  zero. For this case Majorana modes are on the edge of the sample. As a result the effective  sound action $S^{top-sound}$ allows to identify the Topological Superconductor.  Due  too the high order derivatives it is difficult to  observe such  a term in  the laboratory. For this reason we will look for an alternative way to identify the Topological Superconductor.
This result is similar to the  gravitational Chern-Simons term \cite{Witten}. 

In the presence of dislocations we expect to have additional Majorana zero modes. Therefore the sound waves can couple to the Majorana fermions and electron field.
In order to study this effect we will consider small deformations $|\vec{u}(\vec{r},t)|<<a$  ($a$ is the lattice constant). We  will  keep only   first order terms  in  the  coordinate  transformation and  neglect  the effect on the  metric tensor  and  spin connection $\mathbf{\omega}_{i}^{a,b}$.
\begin{eqnarray}
&&H^{(deformed-p-wave)}\approx \frac{1}{2} \int\,d^2r \Big[ C^{\dagger}(\vec{r})\Big(\frac{- \hbar^2}{2m}\delta_{i,j}  \partial_{i} \partial_{j} -\mu(\vec{r})\Big)C(\vec{r})
 -\Delta(\vec{r},t)\Big (\tau^{1}-i\tau^{2}\Big)(E^{i}_{1}\partial_{i}+iE^{i}_{2}\partial_{i})\nonumber\\&&+\Delta^{*}(\vec{r},t) \Big(\tau^{1}+i\tau^{2}\Big)(E^{i}_{1}\partial_{i}-iE^{i}_{2}\partial_{i}) \Big] C(\vec{r},t)\nonumber\\&&
\end{eqnarray}
We define $\delta H^{(cr-p-wave)}$:
\begin{equation}
\delta H^{(cr-p-wave)}\equiv  H^{(deformed-p-wave)}-H^{(p-wave)}
\label{ha}
\end{equation}
The  Hamiltonian $\delta H^{(cr-p-wave)}$  is given  in Eq.$(3)$.

\vspace{0.2 in}

\textbf{Appendix-C}

\vspace{0.2 in}

 \noindent We use  Wick's theorem for  the Green's functions, 
 $G(\vec{k},\tau)=- i\langle G,\epsilon_{a=1}...\epsilon_{a=n}|T(\eta(\vec{k},\tau)\eta^{\dagger}(\vec{k},0))|G,\epsilon_{a=1}...\epsilon_{a=n}\rangle$ for the Superconductor and $g_{a}(\tau)=- i\langle G,\epsilon_{a=1}...\epsilon_{a=n}|T(\zeta_{a}(t)\zeta^{\dagger}_{a}(0))|G,\epsilon_{a=1}...\epsilon_{a=n}\rangle$  for he Majorana fermions. The spinors $\Phi(\vec{k})$, $\hat{W}_{a}(\vec{r},\phi)$   are used to compute the "coherent" factors of the  self energy  $ \mathbf{\Pi}[\vec{q},\omega]$:
\begin{eqnarray}
&&\mathbf{\Pi}[\vec{q},\omega]= i\frac{ \hbar \omega^2 k^{3}_{F}}{8\pi}\Big[1- 2|\Delta|\frac{|\vec{q}|}{|\omega|}\cos[\alpha-\beta(\vec{q})]-4|\Delta|^2(\frac{|\vec{q}|}{|\omega|})^2\Big] \int\,\frac{d\hat{E}}{\sqrt{\hat{E}^2-\Delta^2 k^{2}_{F}}}\cdot\nonumber\\&&
 \frac{(\sqrt{\hat{E}^2-\Delta^2 k^{2}_{F}}-\hat{E})^2}{(\sqrt{\hat{E}^2-\Delta^2 k^{2}_{F}}-\hat{E})^2+\Delta^2 k^{2}_{F}}\sum_{a=1}^{n}\int\,d \tau\Big[\Big(\langle G,\epsilon_{a=1}...\epsilon_{a=n}|T\Big(\eta^{\dagger}(\hat{E},\tau)\eta(\hat{E},0)\Big)|G,\epsilon_{a=1}...\epsilon_{a=n}\rangle\nonumber\\&& +\langle G,\epsilon_{a=1}...\epsilon_{a=n}|T\Big(\eta(\hat{E},\tau)\eta ^{\dagger}(\hat{E},0)\Big)|G,\epsilon_{a=1}...\epsilon_{a=n}\rangle\Big )\Big(\langle G,\epsilon_{a=1}...\epsilon_{a=n}|T\Big(\zeta^{\dagger}_{a}(\tau)\zeta_{a}(0)\Big)|G,\epsilon_{a=1}...\epsilon_{a=n}\rangle\nonumber\\&& +\langle G,\epsilon_{a=1}...\epsilon_{a=n}|T\Big(\zeta_{a}(\tau)\zeta ^{\dagger}_{a}(0)\Big)|G,\epsilon_{a=1}...\epsilon_{a=n}\rangle \Big)\Big]\nonumber\\&&
\end{eqnarray}
$\tau=t_{1}-t_{2}$, $T$ stands for  the time order product. In Eq.$(17)$ we have used the notation  $\hat{E}=\frac{E}{\hbar v}$; $\hat{\epsilon}_{a}=\frac{\epsilon_{a}}{\hbar v}$; $tan[\beta (\vec{q})]=\frac{q_{2}}{q_{1}}$.
We integrate with respect   $\tau=t_{1}-t_{2}$  and  obtain:
\begin{eqnarray}
&&\mathbf{\Pi}[\vec{q},\omega]= \frac{ \hbar \omega^2 k^{3}_{F}}{8\pi}\Big[1- 2|\Delta|\frac{|\vec{q}|}{|\omega|}\cos[\alpha-\beta(\vec{q})]-4|\Delta|^2(\frac{|\vec{q}|}{|\omega|})^2\Big] \int\,\frac{d\hat{E}}{\sqrt{\hat{E}^2-\Delta^2 k^{2}_{F}}}\cdot\nonumber\\&&
 \frac{(\sqrt{\hat{E}^2-\Delta^2 k^{2}_{F}}-\hat{E})^2}{(\sqrt{\hat{E}^2-\Delta^2 k^{2}_{F}}-\hat{E})^2+\Delta^2 k^{2}_{F}}\sum_{a=1}^{n}\Big[\frac{\Theta[\hat{E}]\Theta[\hat{\epsilon}_{a}]}{\omega +(\hat{E}+\hat{\epsilon}_{a})-ix}+\frac{\Theta[\hat{E}]\Theta[\hat{\epsilon}_{a}]}{\omega -(\hat{E}+\hat{\epsilon}_{a})+ix}\nonumber\\&&+\frac{\Theta[-\hat{E}]\Theta[-\hat{\epsilon}_{a}]}{\omega -(\hat{E}+\hat{\epsilon}_{a})-ix}+\frac{\Theta[-\hat{E}]\Theta[-\hat{\epsilon}_{a}]}{\omega +(\hat{E}+\hat{\epsilon}_{a})+ix} +\frac{\Theta[-\hat{E}]\Theta[\hat{\epsilon}_{a}]}{\omega +(\hat{E}-\hat{\epsilon}_{a})-ix}+\frac{\Theta[-\hat{E}]\Theta[\hat{\epsilon}_{a}]}{\omega-(\hat{E}-\hat{\epsilon}_{a})+ix}\nonumber\\&&+
\frac{\Theta[\hat{E}]\Theta[-\hat{\epsilon}_{a}]}{\omega +(\hat{E}-\hat{\epsilon}_{a})-ix}+\frac{\Theta[\hat{E}]\Theta[-\hat{\epsilon}_{a}]}{\omega -(\hat{E}-\hat{\epsilon}_{a})+ix}\Big],x\rightarrow 0\nonumber\\&&
\end{eqnarray}
 At finite temperatures the steps functions are replaced by the Fermi Dirac ocupation functions,  $\Theta[-\hat{E}]=n_{F}[\hat{E}]=\frac{1}{e^{\beta \hat{E}}+1}$,   $\Theta[-\hat{\epsilon}_{a}]=n_{F}[\hat{\epsilon}_{a}]=\frac{1}{e^{\beta \hat{\epsilon}_{a}}+1}$,$\Theta[\hat{E}]=1-\Theta[-\hat{E}]$,  $\Theta[\hat{\epsilon}_{a}]=1-\Theta[-\hat{\epsilon}_{a}]$.


\end{document}